# Table-top Nonlinear Extreme Ultraviolet Spectroscopy

T. Helk[1,2,*,†], E. Berger[3,4,†], S. Jamnuch[5,†], L. Hoffmann[6], A. Kabacinski[7], J. Gautier[7], F. Tissandier[7], J. P. Goddet[7], H.-T. Chang[3], J. Oh[3], C. D. Pemmaraju[8], T. A. Pascal[5,9,10], S. Sebban[7], C. Spielmann[1,2,*] & M. Zuerch[3,4,6,*]

[1] *Institute of Optics and Quantum Electronics, Abbe Center of Photonics, Friedrich-Schiller University, 07743 Jena, Germany*
[2] *Helmholtz Institute Jena, 07743 Jena, Germany*
[3] *Department of Chemistry, University of California, Berkeley, CA 94720, USA*
[4] *Materials Science Division, Lawrence Berkeley National Laboratory, Berkeley, CA 94720, USA*
[5] *ATLAS Materials Physics Laboratory, Department of Nano Engineering and Chemical Engineering, University of California San Diego, La Jolla, CA 92023, USA*
[6] *Fritz Haber Institute of the Max Planck Society, 14195 Berlin, Germany*
[7] *Laboratoire d'Optique Appliquee, ENSTA Paris, Ecole Polytechnique, CNRS, Institut Polytechnique de Paris, Palaiseau, France*
[8] *Stanford Institute for Materials and Energy Sciences, SLAC National Accelerator Laboratory, Stanford, CA 94025, USA*
[9] *Materials Science and Engineering, University of California San Diego, La Jolla, CA, 92023, USA*
[10] *Sustainable Power and Energy Center, University of California San Diego, La Jolla, CA, 92023, USA*

[†]These authors contributed equally.

**The lack of available table-top extreme ultraviolet (XUV) sources with high enough fluxes and coherence properties have limited the availability of nonlinear XUV and x-ray spectroscopies to free electron lasers (FEL). Here, we demonstrate second harmonic generation (SHG) on a table-top XUV source for the first time by observing SHG at the Ti $M_{2,3}$-edge with a high harmonic seeded soft x-ray laser (HHG-SXRL).[1,2] Further, this experiment represents the first SHG experiment in the XUV. First-principles electronic structure calculations are used to confirm the surface specificity and resonant enhancement of the SHG signal. The realization of XUV-SHG on a table-top source with femtosecond temporal resolution opens up tremendous opportunities for the study of element-specific dynamics in multi-component systems where surface, interfacial, and bulk-phase asymmetries play a driving role in smaller-scale labs as opposed to FELs.**

Nonlinear interactions between light and matter form the basis for the generation of light at wavelengths spanning the THz to x-ray regimes and also enable spectroscopies that yield unique insight into fundamental material properties.[3–5] When describing the nonlinear light-matter interaction, a material's polarization response to incident light of frequency ω can be approximated as a power series in increasing powers of the electric field E(ω), where only non-centrosymmetric materials, interfaces, and surfaces permit nonvanishing even-order terms. For this reason, SHG and sum-frequency generation (SFG) are inherently sensitive to spatial asymmetries.[6,7] The nonlinear susceptibility can be extracted experimentally via the relationship

$$I(2\omega) \propto |\chi^{(2)}(2\omega; \omega + \omega)|^2 I(\omega)^2 \quad (1)$$

where I(ω) is the intensity at frequency ω and $\chi^{(2)}$ is the effective second-order susceptibility. While optical nonlinear spectroscopies have been highly insightful probes of interfacial chemistry[8–10], optical light couples efficiently to multiple excitation pathways rendering spectra difficult to interpret in multi-component systems.[11,12] The desire to gain core-level specificity in nonlinear spectroscopies is thus motivated by the study of complex systems with wide-ranging applications including all-solid-state batteries with multiple buried interfaces, ferroelectric materials that, by definition, have bulk-phase structural asymmetries, and low-dimensional heterostructures, to name a few.

Popular element-specific probing methods include photoelectron and x-ray spectroscopies, which are either sensitive to surface- or bulk-properties, respectively, but not both, prompting recent efforts to apply nonlinear spectroscopies to the x-ray regime.[13] The first promising experiments at FELs have demonstrated soft x-ray (SXR) SHG as a viable technique to study non-centrosymmetric materials[14,15], surfaces[16], and buried organic-inorganic interfaces[17] when the energy of the fundamental or its SHG counterpart matches an allowed electronic transition. With FEL sources, the inherent intensity jitter lends itself naturally to calculating the nonlinear response via equation (1). While progress has been made in SHG studies at large-scale x-ray facilities, the next frontier lies in the development of compact table-top sources capable of delivering high enough x-ray photon fluxes to enable nonlinear spectroscopic experiments. Given the moderate input intensities (~$10^{12}$ W/cm$^2$) required for FEL-SHG, it is not unreasonable to think that high-power table-top HHG sources could be used in a similar scheme.[1,16,18–20] In addition, while SHG has been demonstrated with SXR and optical light, nonlinear spectroscopy has been unreported in the XUV regime with a table-top setup. Here, we report the first XUV-SHG experiment above the Ti M-edge (32.6 eV), which also represents the first table-top demonstration SHG using a SXR source. A comparison of our observations with density functional perturbation theory (DFPT) and real-time time-dependent density functional theory (RT-TDDFT) calculations reveals resonantly-enhanced SHG from Ti-3p to Ti-3d states at the Ti surface. The presented scheme therefore opens a pathway for table-top XUV/SXR nonlinear spectroscopy as a sensitive probe of material properties.

**Results**
In the experimental setup depicted in Fig. 1a, a 37.8 eV HHG-SXRL[1,2] with an average input energy of 111 ± 23 nJ, pulse duration of 1.73 ± 0.13 ps, and Gaussian-like beam profile (Fig. 1a, inset) was tightly focused using a Au-coated ellipsoidal mirror onto a Ti foil (hexagonal centrosymmetric, 6/mmm point group) placed at the rear focal plane of the ellipsoid with an on-target spot size of 4.5 ± 1.5 μm, and average intensity of 4.1 ± 1.9 x $10^{11}$ W/cm$^2$. Given the 50 nm thickness of the Ti foil and the 13 nm attenuation depth of 37.8 eV photons in Ti, it can be concluded that the SHG signal was predominately generated on the front face of the foil (Fig. 1b). The on-target fluence of $0.7 \pm 0.3$ J/cm$^2$ surpassed the single shot damage threshold as evidenced by consistent sample damage, which required moving the sample to an unexposed spot for each laser shot. Since the Ti conduction band (CB) consists primarily of 3d states, the 37.8 eV linear absorption can be attributed to a resonant dipole-allowed inter-shell transition from 3p core states to empty 3d CB states. The SHG emerges from a subsequent transition to a virtual state 43 eV above the Fermi energy $E_F$ (Fig. 1c).

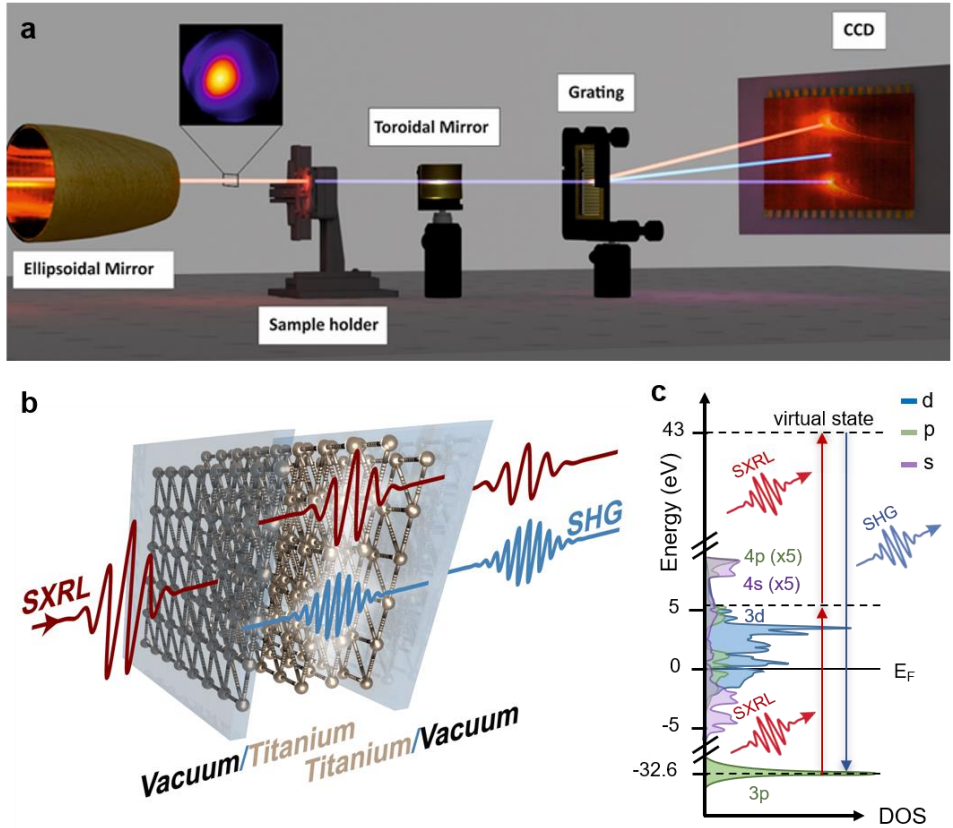

**Fig.1: Experimental setup and energy diagram for SXRL-SHG at the Ti M$_{2,3}$ edge. a)** From left, the setup includes the incoming 37.8 eV HHG-SXRL pulse (red beam, far-field beam profile in inset) that is focused by an ellipsoidal mirror onto the Ti foil surface, where SHG is generated (blue beam) and propagates collinearly with the fundamental. The divergent field is refocused with a toroidal mirror onto a grating that disperses the SHG and fundamental beams simultaneously onto a CCD camera. **b)** Schematic diagram of the origin of the SHG emission. Inversion symmetry is broken on the front Ti surface, allowing for SHG. The fundamental and SHG beams exit from the rear foil surface. **c)** Ti Orbital-resolved density of states.[30] The SHG emission results from on-resonant excitation of 3p core states (-32.6 eV) to an empty intermediate state of 3d character (+5.2 eV) and subsequently to a virtual state (+43 eV).

The outgoing photons were refocused using a Au toroidal mirror and dispersed with a transmission grating (1000 lines/mm) onto a deep-cooled charged coupled device (CCD,

Princeton-MTE), enabling a simultaneous analysis of the fundamental and SHG peaks. The pulse energy was calibrated with respect to the counts on the CCD by measuring the shot-to-shot fluctuations of the SXRL and correlating the statistics with those observed on the CCD. A characteristic spectrum featuring the fundamental and SHG peaks is shown in Fig. 2a. Measurements of the seeded SXRL linewidth using a high-resolution spectrometer confirmed a linewidth at FWHM of 2.6 meV, which is poorly resolved, but expected since the spectrometer was optimized to cover more than one octave at low resolution to be sensitive for SHG. The broadened linewidth of the second harmonic is due to aberrations of the imaging spectrometer, but otherwise expected to be the same order of magnitude as that of the fundamental. The background-corrected spectra were analysed by integrating the peaks observed on the CCD and aggregating single-shot spectra according to the intensity of the fundamental. The slope of a plot relating the on-target energies of the fundamental and SHG peaks allows retrieval of $\chi^{(2)}$ via equation (1) (Fig. 2b).

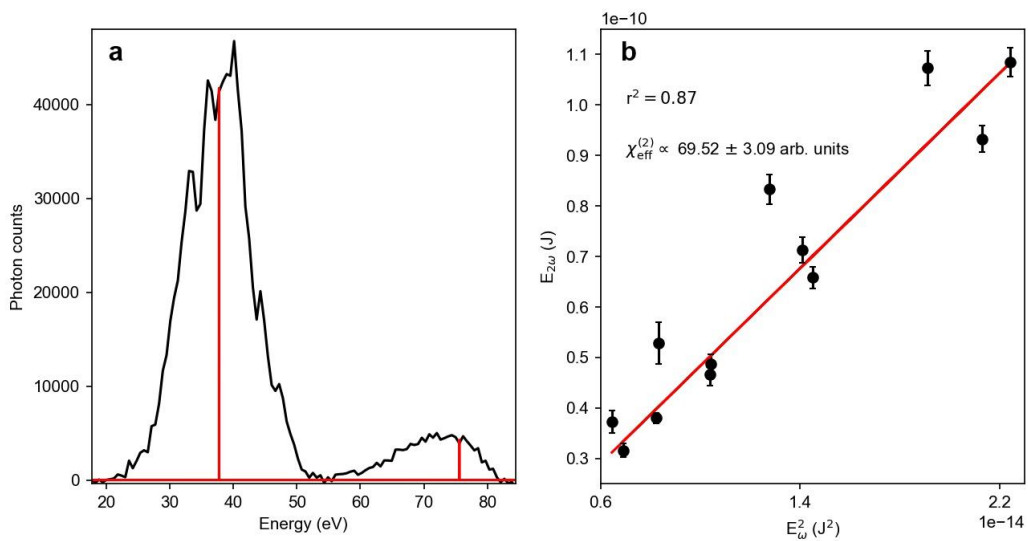

**Fig. 2: Second harmonic generation at the titanium $M_{2,3}$ edge. a)** A representative spectrum from of the fundamental and SHG signals (black). The linewidth is not resolved in this spectrum due to the spectrometer being optimized for high sensitivity exhibiting poor spectral resolution. The true line width of the SXRL was separately measured ($\Delta E \approx 2.6$ meV) with a high resolution spectrometer (red line). The linewidth of the SHG signal is assumed to be of the same order of magnitude as that of the fundamental. **b)** The nonlinear energy dependence of the SHG signal with respect to the fundamental. A linear equation (red line) is fit to experimental data plotted as $|E(\omega)|^2$ vs. $E(2\omega)$ (black dots with one $\sigma$ error bars, determined by the standard deviation in photon counts on the CCD resulting from averaging shots at the same fundamental pulse energy together).

To understand the SHG process in detail, first principles DFPT[21] calculations were performed using the *exciting* all-electron full-potential computer package employing the linearized augmented planewave and local orbital methods.[22,23] Following the formalism of Sharma[24], we calculated the expected frequency domain SHG signal from the bulk and surface Ti atoms. We applied a 3.2 eV zero-point energy shift to the calculated spectrum, based on comparison of the simulated linear response spectrum to our measurements. The SHG response of the Ti shown in Fig. 3a depicts the calculated nonlinear susceptibility across the Ti $M_{2,3}$ edge primarily originating from surface Ti atoms. Next, RT-TDDFT calculations of the Ti slab were performed using a development version of the SIESTA[25] electronic structure code. The Ti slabs were driven under an XUV monochromatic pulse to investigate their interaction with varying laser field intensity. The nonlinear second harmonic response was confirmed and the second order susceptibility calculated by fitting J(ω) to a quadratic equation in the driving electric field strength (see SI). In our experiments, the incident photon energy fell within this range (dashed gray line, Fig. 3a), as confirmed by the agreement between the experimentally measured and theoretically calculated linear absorption spectrum (Fig. 3b), thus allowing efficient SHG.

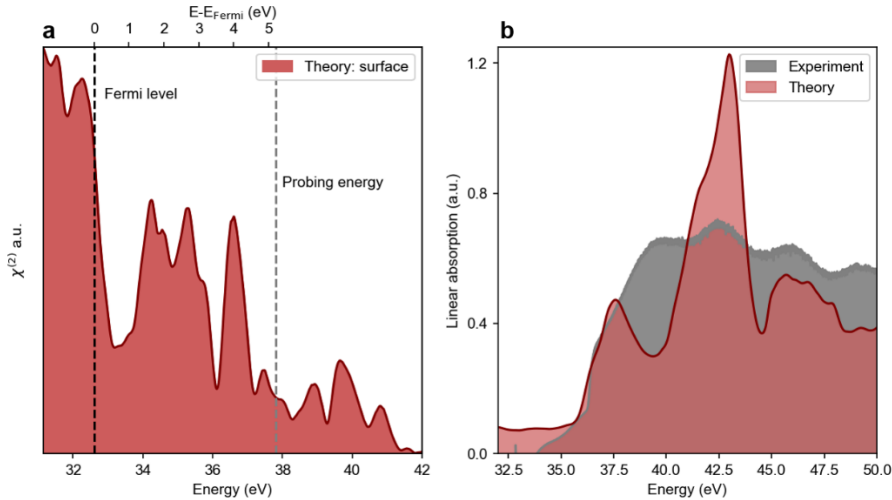

**Fig. 3: Realtime TDDFT calculation for $\chi^{(2)}$ surface spectrum and comparison between theory and experiment. a)** The DFT calculated $\chi^{(2)}$ spectrum originating from the Ti bulk-atoms the Ti surface-atoms (red). The bulk contribution is negligible due to symmetry considerations. The probing energy used in this experiment (37.8 eV) is overlaid (grey dashed line). **b)** Comparison of the linear absorption of Ti obtained experimentally (gray) from a broadband HHG source and calculated from DFT (red).

**Discussion**

We find that the calculated $\chi^{(2)}(\omega)$ spectrum is in good agreement with experiments with the overwhelming number of 3d states just above the Fermi level explaining the $\chi^{(2)}$ spectrum peaking in this region. The bulk response, on the other hand, is negligible due to the centrosymmetry of the Ti unit cell and its numerical value is suppressed by more than seven orders of magnitude.

The quadratic modulo intensity dependence of the second-harmonic signal at 75.6 eV with respect to the energy of the fundamental (37.8 eV) was confirmed by an $R^2$ value of 0.87 (Fig. 2b) when plotting $E_\omega^2$ vs $E_{2\omega}$. Using the mean yield of the SHG, fundamental, and input parameters of the SXRL, the SHG conversion efficiency was estimated to be $\eta = 2.3 \pm 0.5 \times 10^{-2}$. Given SXRL incident energies in the nJ range, this high conversion efficiency for a nonlinear process is remarkable. Previous FEL experiments estimated a threshold power for nonlinear effects in the hard x-ray (HXR) regime exceeding $10^{16}$ W/cm$^2$.[14] In the SXR regime, highly efficient SHG at incident powers of $\approx 10^{12}$ W/cm$^2$ were observed but with μJ input photon energies.[16,26] Our results demonstrate efficient SHG with noticeably less power ($\approx 10^{11}$ W/cm$^2$), which could stem from the absorption edges involved in the process. For example, previous SXR-SHG experiments probing C and B K-edges might result in a comparatively smaller enhancement than at the Ti M-edge, merely due to the fewer number of K-shell electronic states and the weaker coupling of dipole moments between K-shell and valence electrons.[16,17] Although beyond the scope of this first experimental work in the XUV, it appears worthwhile to explore how the nonlinear conversion efficiencies scale with the energies of core states in future theoretical work.

In conclusion, we have shown the first successful demonstration of XUV-SHG with surface-specificity confirmed by first-principles calculations and performed nonlinear spectroscopy on a table-top soft x-ray laser system for the first time. In contrast with the energies required for SHG at FEL sources, it was possible that a table-top SXRL with nJ input energies and tight-focusing could generate second harmonic radiation at 75.6 eV. Additionally, the large number of transition metal elements with M-edges in the XUV window make XUV-SHG a feasible technique to investigating intrinsic properties of transition metal molecular complexes, heterojunctions, interfaces, and non-centrosymmetric materials with elemental-specificity. The findings presented here hint at new opportunities to perform time-resolved femtosecond or even attosecond nonlinear XUV spectroscopy experiments using, for example, the high energy XUV-HHG-based sources at the Extreme Light Infrastructure[27] or up-scalable sources based on relativistic surface harmonics.[28] Our first demonstration of nonlinear XUV spectroscopy on the table-top holds great promise to expand the nonlinear suite to tunable XUV-SFG to record broadband surface and interface spectra and reduce the temporal resolution by using few-femtosecond compressed optical gated pulses.

## *Materials and Methods*

### **Experimental Design**

The goal of the experiment was to measure SHG from a Ti surface at a SXRL. The linear XUV absorption spectrum was first measured with a table-top HHG source. A sub-4 fs, broadband NIR pulse centered at 730 nm was focused into an Ar gas jet, producing high-harmonics. The linear absorption spectrum was collected by measuring the broadband XUV transmission through vacuum ($T_{vac}$) and through a 50 nm thick Ti foil ($T_{Ti}$). Sets of these two measurements were repeated 127 times to determine the average absorbance A = -log($T_{Ti}/T_{vac}$). The SHG measurements were subsequently conducted at the Laboratoire d'Optique Appliquée using the Salle Jaune Ti:Sapphire laser system[2], able to deliver three independently-compressed multi-terawatt femtosecond pulses at a repetition rate of 1 Hz. The SXRL works by coupling a resonant HHG pulse into an SXR amplifier. The amplifier is a plasma of $Kr^{8+}$ ions emitting at the $3d^94d_{(J=0)}$ → $3d^94p_{(J=1)}$ transition at 32.8 nm (37.8eV) generated by optical-field ionization of a high-density Kr gas jet by an ultrashort infrared pump pulse (1.5J, 30fs, focused at $3 \times 10^{18}$ W/cm$^2$). Due to the high electron density of the plasma (up to $10^{20}$ e$^-$/cm$^3$), the pump pulse cannot propagate in the plasma. A waveguide was therefore implemented beforehand by focusing a 0.7 J, 0.6 ns infrared pulse in the gas jet using an axicon lens.[29] The HHG was seeded with a 15mJ, 30fs infrared pulse focused in an Ar-filled gas cell and coupled into the $Kr^{8+}$ amplifier using a grazing-incidence toroidal mirror. The 25$^{th}$ harmonic of the HHG spectrum was tuned to the lasing transition by chirping the driver pulse, and the injection delay was set to 1.2ps to match the SXRL gain peak position.

The SXRL was focused with an ellipsoidal mirror of a focal length of 33 cm (3 µm diffraction limit), onto the sample. An upper limit on the focal spot of 6 µm was estimated by the distance it required to move the position of the incident beam in the sample plane between laser shots due to sample damage. The irradiated spot size was thus estimated be 4.5 ± 1.5 µm. To ensure the sample plane was aligned with the focal position, the z-position (optical axis) of the sample was varied until a threshold intensity strong enough to burn holes into the foil were reached. After hitting the sample, the fundamental and SHG beams were focused with a toroidal mirror (f = 33 cm) onto a cooled CCD camera (1024 x 1024 pixels, pixel-pitch 13.5 µm, -50 °C). A total of 394 spectra were recorded, 248 of which had full resolution of the CCD-chip and 146 of which with a hardware binned camera of 2 x 2 pixels.

**Acknowledgements**

The research leading to these results has received funding from the European Community's Horizon 2020 research and innovation program under grant agreement No. 654148 (Laserlab Europe). M.Z. acknowledges support by the Max Planck Society (Max Planck Research Group) and the Federal Ministry of Education and Research (BMBF) under "Make our Planet Great Again - German Research Initiative" (Grant No. 57427209 "QUESTforENERGY") implemented by DAAD. This work is supported by Investissements dAvenir Labex PALM (ANR-10-LABX-0039-PALM). The authors thank Dr. C. Schwartz, C. Uzundal and F. Tuitje for fruitful discussions. J.O. is supported by Basic Science Research Program through the National Research Foundation of Korea funded by the Ministry of Education (2019R1A6A3A03032979). H.-T. C. acknowledges support from the Air Force Office of Scientific Research (No. FA9550-14-1-0154) and the W.M. Keck Foundation (No. 046300). The XUV transient absorption experiment is funded by the Air Force Office of Scientific Research (No. FA9550-14-1-0154) and the Army Research Office (No. W911NF-14-1-0383). This research used resources of the National Energy Research Scientific Computing Center, a DOE Office of Science User Facility supported by the Office of Science of the U.S. Department of Energy under Contract No. DE-AC02-05CH11231.

**Author Contributions**
T.H., L.H., A.K., J.G., F.T., J.P.G., S.S. and M.Z. performed the experiments. T.H., E.B., and M. Z. analysed the data. S. J., C.D.P. and T.A.P. performed the simulations. J.O. and H.T.C. performed linear absorption measurements. M. Z. conceived the experiment. C. S. and M. Z. supervised the project. All authors contributed to the manuscript.

**Corresponding Authors:** Correspondence and requests for materials should be addressed to T. Helk (tobias.helk@uni-jena.de), C. Spielmann (christian.spielmann@uni-jena.de) and M. Zuerch (mwz@berkeley.edu).

**Competing Interests**
The authors declare no competing financial interests.

# Supplementary Materials

**Processing Experimental Data**

Figure S1a depicts a logarithmic image of 146 accumulated single shots of a binned (512 x 512 pixel) camera image. A non-trivial background that increases vertically is observed, which is particularly strong above pixel 40 on the angular axis. Additionally, diffraction spots above the $0^{th}$ and $\pm 1^{st}$ order peaks along the vertical axis were observed in the raw images. Both of these effects were challenging to remove during the experiment, since a background correction without the sample in the beam path would have been misleading due to the intensity jitter of the SXRL. A post-processing algorithm was implemented to correct for the background that involved manually selecting a sufficient amount of *n* pixels on the image corresponding to the background, fitting a two-variable polynomial of power *n* to the selected pixels, and subtracting the result from each image. A logarithmic image of 248 accumulated and background-corrected single shots of an un-binned camera image (1024 x 1024 pixels) is shown in Figure S1b, with red boxes indicating the regions of analysis. The $\pm 1^{st}$ order peaks are displayed to the left and right of the central peak. The larger peaks on the far right and left correspond to the incident fundamental beam, whereas the smaller peaks second from the left and right correspond to the SHG peaks. Accumulating the 248 raw images was necessary to visualize the SHG peak, as this peak was not apparent in single shot images (Figure S2a – raw image, Figure S2b – background corrected).

To extract the $\chi^{(2)}$ response according to equation (1) in the main text, the spectra were sorted into bins based on the number of photon counts within the areas corresponding to the $\pm 1^{st}$ order of the fundamental. Within each bin, the spectra were accumulated and normalized, and the photon counts within areas corresponding to the $\pm 1^{st}$ order of the SHG peaks were obtained. The error of the fundamental and SHG yield was calculated assuming Gaussian distribution within each bin and determining the one $\sigma$ width. For the binned data set, the SHG intensity is on the order of the un-binned data, but with a less intensive fundamental beam. This can be explained by the numerical background correction, since binning increases the yield for each pixel. The simulated background is therefore also higher and reduces the input yield of the fundamental, which makes it harder to compare the both data sets. The square of the pulse energies obtained for the fundamental and that of the SHG peaks were plotted with respect to each other. A linear equation was fit to the data with $R^2=0.87$. The dependence of the SHG-yield is offset corrected with the y-intercept from the fit function to ensure that zero input intensity of the fundamental corresponds to zero SHG output. A negative yield in this graphs would be caused by this offset subtraction.

The conversion of photon counts on the CCD to on-target pulse energy was performed by measuring the shot-to-shot statistics of the SXRL with a footprint camera in place of the experimental setup and correlating the fluctuations with those observed on the CCD camera. For measuring the pulse energy with the footprint camera, two 0.45 µm thick Al filters (transmission ≈ 2%) and a reflective mirror with an angle of incidence of 45° (reflectivity ≈ 30%) were used to attenuate and steer the SXRL beam onto the camera (quantum efficiency at 37.8 eV ≈ 70%). The beam characterization camera setup was then replaced with one 0.15 µm thick Al filter (transmission ≈ 50%) and an Au ellipsoidal mirror (reflectivity ≈ 70%) and the remainder of the experimental setup as seen in Figure 1a. From this, the average energy of the SXRL was determined to be 111 ± 23 nJ. The pulse energies of the SHG were determined by considering the different optical losses from the beam path taken by the SHG photons from sample to CCD camera (Table S1).

**First-Principles Simulation of Linear and Nonlinear Susceptibility**

We first assessed the frequency-dependent susceptibility $\chi^{(2)}$ of the Ti surface second harmonic response by means of density functional perturbation theory (DFPT) calculations using the *exciting*[22] all-electron software package. Structures representing the Ti bulk and slab, constistening of eight atoms were used to simulate the second harmonic response. The Ti system was modeled as a slab supercell of 6 layers of hexagonal close-packed unit cells, with dimensions a = 2.95 Å and c = 17.5725 Å along with 12.5 Å of vacuum. Here, the formalism by Sharma[24] implemented within *exciting* was used to determine the second-order response. A total of 200 empty states were included in the ground state calculation to account for the excited states, which was deemed enough to increase the number of energy eigenvalues to extend above $2\hbar\omega$, while still being computationally feasible. The symmetry in the system was artificially broken by excluding half of the Ti semi-core 3s and 3p electrons, which are the source of the relevant SHG response from the self-consistent field calculation loop. The Brillouin zone was sampled with a 40x40x1 Γ-point centered k-point grid. To extract the nonlinear response, the background signal from the valence electrons was eliminated by fitting the real and imaginary part of $\chi^{(2)}$ at higher energies to be proportional to the reciprocal of the energy.[16] A rigid shift of 3.2 eV was necessary to align the DFT linear response to the experimental spectral features.

**Real-Time TDDFT Simulation of Linear and Nonlinear Response**

The material response of a monochromatic laser field incident on a Ti foil was simulated with real-time velocity gauge[30,31] TDDFT using a linear combination of localized atomic orbitals implemented within SIESTA.[25,32] The linear response of Ti was calculated first as a benchmark for calibrating the laser field energy.[33] Here, we employed a semi-core norm-conserving pseudopotential including Ti 3s and 3p states in the valence band to include the response at the desired Ti-M edge. The exact same geometry from the previous first-principles calculation was used for this part. The Brillouin zone of the slab was sampled with a Γ-centered 20x20x1 k-point grid. Exchange-correlation effects within TDDFT were simulated at the level of the Perdew-Zunger local density approximation (LDA-PZ).[34] We employed a basis set of double-ζ quality consisting of {3s(ζ), 3p(ζ), 4s(2ζ), 3d(2ζ), 4p(ζ)} orbitals for a total of 19 atomic orbitals per Ti atom. The real space mesh cutoff was set to 400 Ry. A timestep of 0.04 a.u. (1.935 as) was used to propagate the system.

The current response J(t) of bulk Ti induced by a weak impulse electric field of 0.001 a.u. along x- and z-axes at time zero is shown in Fig. S3. Due to the symmetry of bulk Ti , material response in the x- and y-directions responded identically to the incident field, whereas a different current response was observed along the z-axis. The frequency-dependent dielectric response function ε(ω) was then obtained from Fourier transformation of J(t). Here, a maximum in the linear response at 41.6 eV was identified (See figure 3(a) of main text). This was used as the external energy of the laser for simulating the nonlinear light-matter interaction.

Next, we investigated the interaction of the Ti slab with a monochromatic laser of field of varying intensity ranging from $1 \times 10^{10}$ to $1 \times 10^{13}$ W/cm$^2$. Under our experimental conditions, we expected an SHG response only at the surface of the Ti as the centrosymmetric bulk extends effectively to infinity. Due to the finite length of the slab in simulations, special care needed to be taken to ensure the SHG response was not predicted from both the front and rear ends of the slab that would cancel each other out and attenuate the SHG signal. Practically, only one Ti surface was made SHG-active by modeling the top half of the supercell with a semi-core electron basis set and the bottom half with a valence pseudopotential that excluded Ti 3s and 3p semi-core states. Further details on this approach can be found in the supplementary information of [16]. Due to the extended nature of highly excited electronic states, ghost atoms were added to the top and bottom of the slab to improve the quality of the non-linear response. These ghost atoms do not impact the position of spectral peaks and were added to account for potential transitions to states in the continuum that would otherwise be absent in our localized orbital basis set approach. The light-matter interaction was then simulated with a 41.6 eV, 5 fs driving pulse with a sine-squared envelope oriented along z-axis perpendicular to Ti slab supercell. Though 5 fs is significantly shorter than the experimental pulse duration, it is sufficiently long to resonantly probe Ti M-edge excitations. This chosen energy of 41.6 eV corresponds to the maximum in the linear spectral response from the TDDFT approach. The slab model was propagated in time with different driving laser fields in the $10^{10}$-$10^{13}$ W/cm$^2$ intensity range (Fig. S4) The incident laser pulse induces a current response J(t) in the Ti slab, whose Fourier transform at twice the driving frequency J(2$\omega$) was used to extract the energy-dependent $\chi^{(2)}$ response.

In Fig. S5, we plot the z-component of the time profile of the laser field and the induced current response of the Ti slab. The corresponding Fourier-transformed current J($\omega$) demonstrates that the second harmonic response at 2$\omega$ is stronger with stronger laser intensity (Fig. S6). The generalized nonlinear susceptibilities can be given by the Taylor-expansion in equation (1) in the main text. This equation is valid in the nonlinear regime and it shows the response at 2$\omega$ scales with the square of the field strength. As shown in Fig S6 we find that J(2$\omega$) exhibits a quadratic dependence with respect to the applied field strength with the quadratic coefficient in the fit being related to $\chi^{(2)}$. Based on this fit, we found the second-order susceptibility $\chi^{(2)}$ to be 0.0206 esu.

| Table S1: Parameters used for determining the on-target intensity of the SXRL beam | |
|---|---|
| Quantum Efficiency of CCD (fundamental) | 0.68 |
| Quantum Efficiency of CCD (SHG) | 0.82 |
| Grating Efficiency (fundamental, SHG) | 0.1 |
| Fraction of beam reflected off toroidal mirror (fundamental) | 0.88 |
| Fraction of beam reflected off toroidal mirror (SHG) | 0.86 |
| Fraction of beam transmitted through Ti foil (fundamental) | 0.019 |
| Fraction of beam transmitted through Ti foil (SHG) | 0.48 |

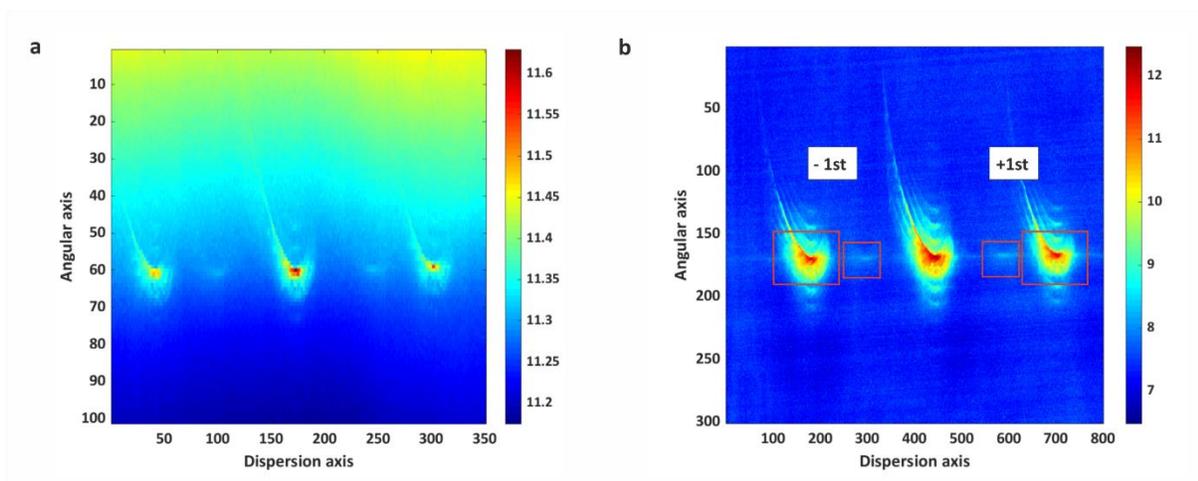

**Fig. S1:** Raw accumulated and background corrected spectra. a) Logarithmic image of 146 accumulated single shots of a binned camera (512 x 512 pixel). A SHG signal is in between zeroth order and ±1st order visible. b) Logarithmic image of 248 accumulated and background corrected single shots of an unbinned camera (1024 x 1024 pixel). Red boxes indicate the area for the analysis of the fundamental and SHG yield of the single spectra.

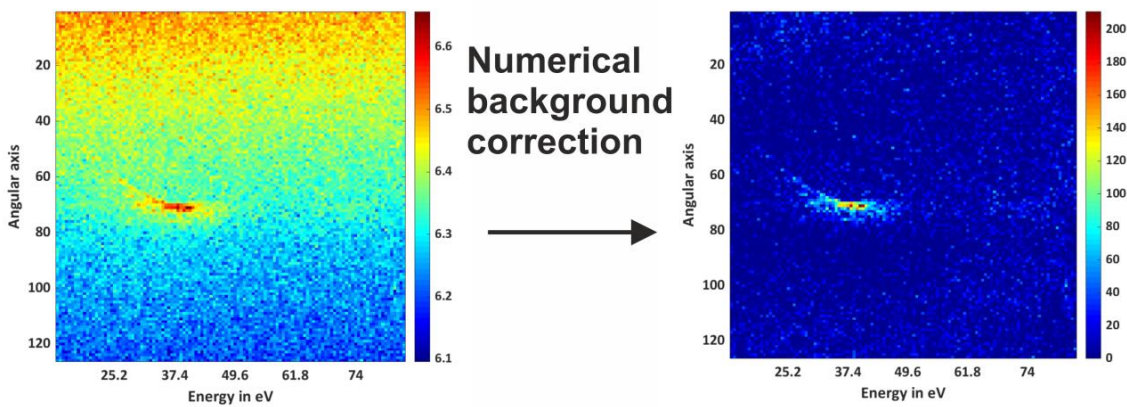

**Fig. S2:** Impact of background correction. On the left, single shot spectra with noisy background in a logarithmic representation. Applying the numerical background correction, described above, leads to spectra as can be seen on the right. A SHG signal is not clearly visible.

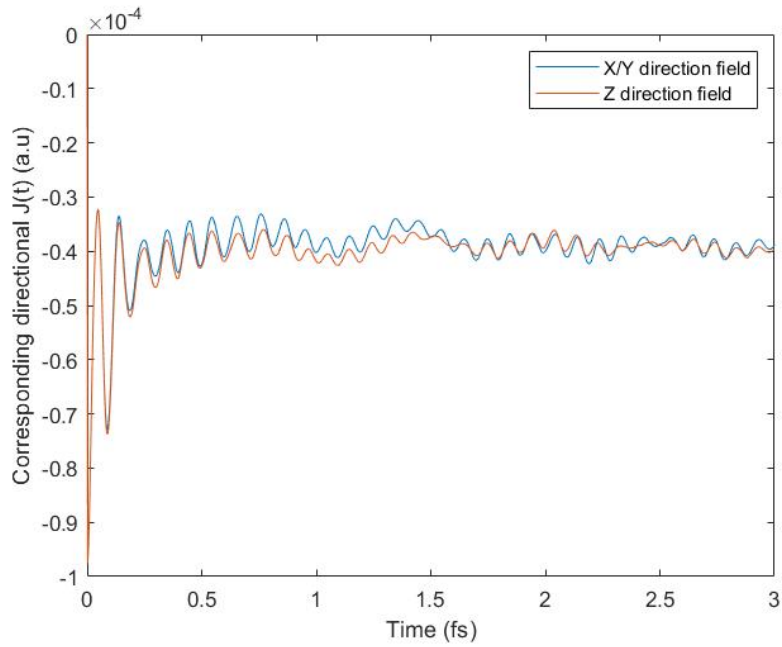

**Fig. S3**: Current response of bulk titanium to a directional impulse perturbation comparison between x/y and z direction.

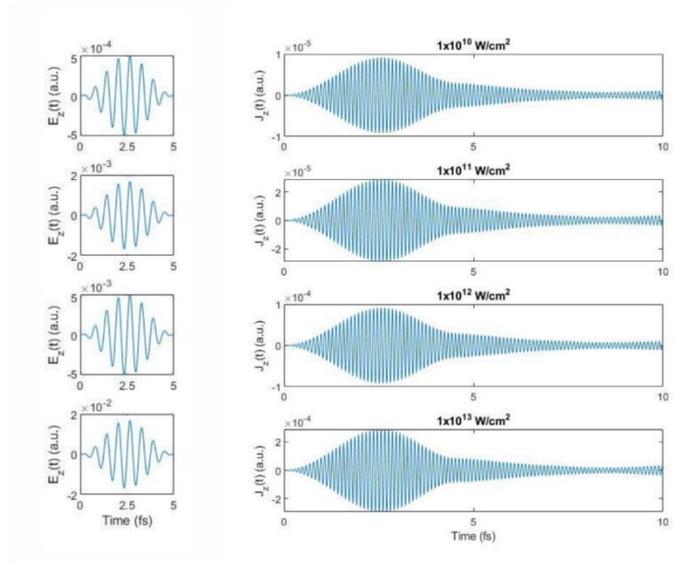

**Fig. S4:** Interaction of laser pulse with titanium slab (left) Magnitude and time profile of applied laser pulse with intensities ranging from $10^{10}$-$10^{13}$ W/cm$^2$ (right) z-component of time dependent induced current response.

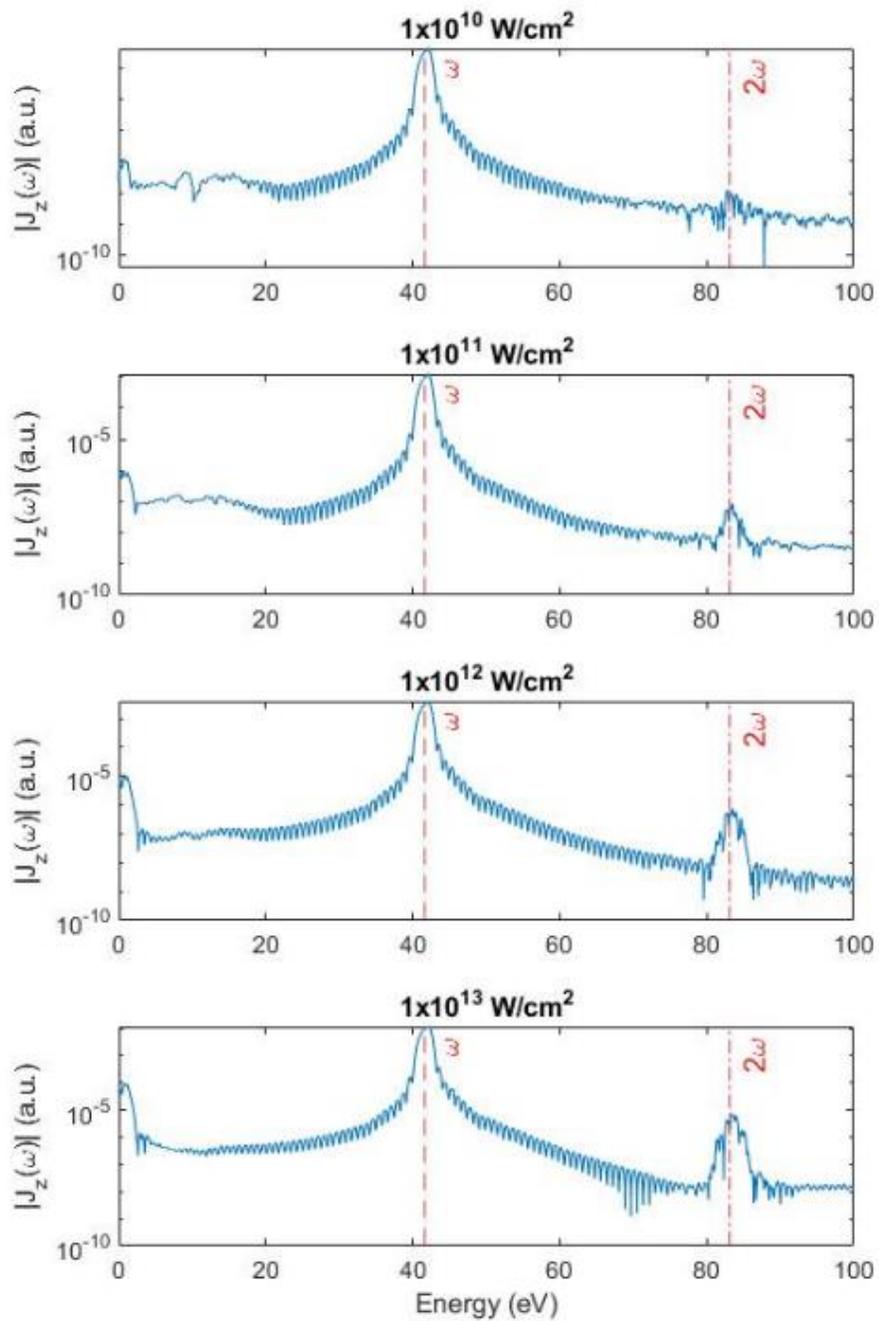

**Fig. S5:** Frequency domain of z-component current function

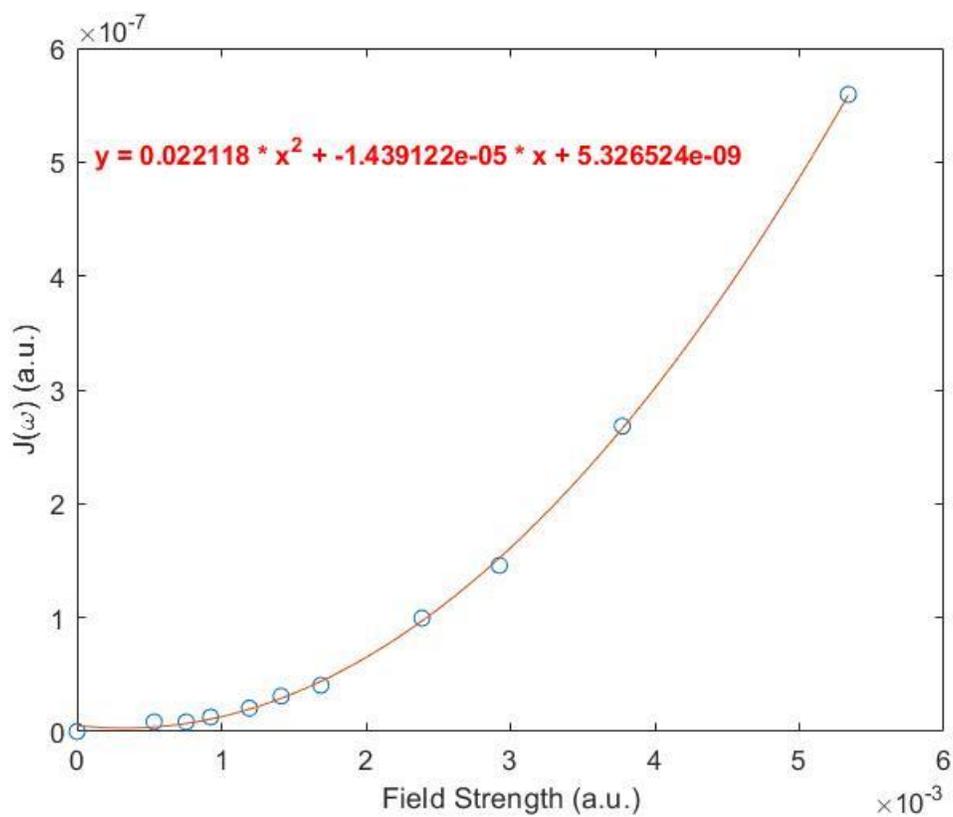

**Fig. S6:** Current response dependence from the input field strength. The fitting to parabolic equation is given above.